\documentclass[referee]{raa}            
\usepackage{graphicx,times}             

\begin{document}

   \title{$R$-mode instability of strange stars and observations of neutron stars in LMXBs
}

   \volnopage{Vol.0 (200x) No.0, 000--000}      
   \setcounter{page}{1}          

   \author{Chun-Mei Pi$^1$
   \and  Shu-Hua Yang$^2$\mailto{}
   \and  Xiao-Ping Zheng$^2$
   }

   \institute{$^1$ Department of Physics and Engineering, Hubei University of Education, Wuhan 430205, China;\\
              $^2$Institute of Astrophysics, Huazhong Normal University, Wuhan 430079, China; \\
            \email{ysh@phy.ccnu.edu.cn}
   }

   \date{Received~~2009 month day; accepted~~2009~~month day}

 \abstract{Using a realistic equation of state (EOS) of strange quark matter, namely, the modified bag model, and considering the constraints to the parameters of EOS by the observational mass limit of neutron stars, we investigate the $r$-mode instability window of strange stars, and find the same result as in the brief study of Haskell, Degenaar and Ho in 2012 that these instability windows are not consistent with the spin frequency and temperature observations of neutron stars in LMXBs.
 }

   \authorrunning{C. M. Pi et al. }            
   \titlerunning{$R$-mode instability of strange stars}  

   \maketitle

%
%
\section{Introduction}

Ever since the realization in 1998 that $r$-modes, which are restored by the
Coriolis force, are subjected to Chandrasekhar-Friedmann-Schutz (CFS) instability (Chandrasekhar 1970; Friedman \& Schutz 1978) in a perfect fluid star with arbitrary rotation (Andersson 1998; Friedman \& Morsink 1998),
its study has received a lot of attention. It is easy to understand that for a realistic neutron star, the $r$-mode instability only happens in
a range of spin frequencies and temperatures, the so-called $r$-mode instability window, which is decided by the competition between the
gravitational-wave driven effect and viscous-dissipation damping effect to the modes (Lindblom et al. 1998). Therefore, primarily the $r$-mode instability is an important physical mechanism that can prevent neutron stars from spinning up to its Kepler frequency ($\Omega_{K}$, above which matter is ejected from the star's equator) (Madsen 1998; Andersson et al. 1999), and gravitational waves emitted during the instability process could be detected (Andersson \& Kokkotas 2001; Andersson et al. 2002; Abadie et al. 2010; Alford \& Schwenzer 2014). In fact, some other aspects related to $r$-mode instability are also studied. For example,
as an alternative explanation to the rapid cooling of neutron star in Cas A (which can be well explained by the superfluidity-triggering model (Page et al 2011; Shternin et al. 2011; Elshamouty et al. 2013)), it is suggested that the star experiences the recovery period following the $r$-mode heating process assuming the star is differentially rotating (Yang et al. 2010; Yang et al. 2011).

Recently, as more and more temperature data of neutron stars                                                                                                                                                                                                                                                                                                                                                                                                                                                     in Low Mass X-ray Binaries (LMXBs) are presented through X-ray and UV observations (Haskell et al. 2012; Gusakov et al. 2014),
many studies are trying to constrain the physics behind the $r$-mode instability of neutron stars, especially the equation of state (EOS) of cold dense matter, by the comparison between the $r$-mode instability window and spin frequency and temperature observations in these systems (Ho et al. 2011; Haskell et al. 2012; Vida\~{n}a 2012; Wen 2012).

In this paper, we will investigate the case of strange stars in detail following the brief study by Haskell et al. (2012). Different from their work and other former works about $r$-modes in strange stars (eg. Madsen 1998; Madsen 2000), our study is based on a realistic EOS of strange quark matter, namely, the modified bag model (Farhi \& Jaffe 1984; Haensel et al. 1986; Alcock et al. 1986; Weber 2005). We give the timescales related to $r$-modes numerically; what's more,
before our study of the $r$-mode instability window, we fix the parameter space of EOS so that it can match the mass limit of neutron stars, which is given by determining the mass of the millisecond pulsar PSR J1614-2230 to be $1.97 \pm 0.04$ $M_{\odot}$ (Demorest et al. 2010),
and has been further updated by the measuring of the $2.01 \pm 0.04$ $M_{\odot}$ PSR J0348+0432 recently (Antoniadis et al. 2013).

Although strange stars can also support a thin crust of normal nuclear matter up to the neutron drip density (Glendenning \& Weber 1992), it only leads to minor changes
 to the maximum mass comparing with bare strange stars (Zdunik 2002), and
it also does not contribute significantly to the damping of $r$-modes (Andersson et al. 2002; Haskell et al. 2012). Therefore, we only study the $r$-mode instability window of bare strange stars in this work, and it will be a very similar one for strange stars with nuclear crust. However, the $r$-mode instability window of strange stars with a crystalline superconducting quark crust will be a very different one as studied by Rupak \& Jaikumar (2013), we will not consider that case in this paper. What's more, we will not consider the case of solid strange quark stars composed by quark clusters (Xu 2003; Yu  \& Xu 2011; Zhou et al. 2014), since $r$-mode instability could not occur in these stars, apparently.

The plan of this paper is as follows. In Section 2, we briefly show the modified strange quark matter EOS taken by our study, and calculate the allowed parameter space following certain constraints. In Section 3, we give the inequality through which the $r$-mode instability window is determined, and the related gravitational-wave driven timescale and the viscous-damping timescales are also presented. In Section 4, we compare the theoretical $r$-mode window with the spin frequency and temperature observations of neutron stars in LMXBs, and Section 5 is our conclusion and discussion.

\section{EOS of strange quark matter and constraints by the mass of PSR J1614-2230 and PSR J0348+0432}

For strange quark matter, we take the modified bag model (Farhi \& Jaffe 1984; Haensel et al. 1986; Alcock et al. 1986; Weber 2005), in which up ($u$)
and down ($d$) quarks are treated as massless particles while
the strange ($s$) quark mass is a free parameter, and first-order perturbative corrections in the strong
interaction coupling constant $\alpha_{S}$ are taken into account. The thermodynamic potential for the $u$, $d$ and $s$ quarks, and for the electrons are (Alcock et al. 1986; Na et al. 2012)

\begin{equation}
\Omega_{u}=-\frac{\mu_{u}^{4}}{4\pi^{2}}(1-\frac{2\alpha_{S}}{\pi}),
\end{equation}

\begin{equation}
\Omega_{d}=-\frac{\mu_{d}^{4}}{4\pi^{2}}(1-\frac{2\alpha_{S}}{\pi}),
\end{equation}

\begin{eqnarray}
\nonumber
\Omega_{s}&=&-\frac{1}{4\pi^{2}}\bigg\{\mu_{s}\sqrt{\mu_{s}^{2}-m_{s}^{2}}(\mu_{s}^{2}-\frac{5}{2}m_{s}^{2})
+\frac{3}{2}m_{s}^{4}f(u_{s},m_{s})-\frac{2\alpha_{S}}{\pi}\bigg[3(\mu_{s}\sqrt{\mu_{s}^{2}-m_{s}^{2}}-m_{s}^{2}f(u_{s},m_{s}))^{2}\\
\nonumber
&&-2(\mu_{s}^{2}-m_{s}^{2})^{2}-3m_{s}^{4}\textrm{ln}^{2}\frac{m_{s}}{\mu_{s}}
+6\textrm{ln}\frac{\sigma}{\mu_{s}}\bigg(\mu_{s}m_{s}^{2}\sqrt{\mu_{s}^{2}-m_{s}^{2}}-m_{s}^{4}f(u_{s},m_{s})\bigg)\bigg]\bigg\},
\label{}
\end{eqnarray}

\begin{equation}
\Omega_{e}=-\frac{\mu_{e}^{4}}{12\pi^{2}},
\end{equation}
where $f(u_{s},m_{s})\equiv\textrm{ln}((\mu_{s}+\sqrt{\mu_{s}^{2}-m_{s}^{2}})/m_{s})$, $\sigma$ is a renomormalization constant whose value is of the order of the chemical potentials (Farhi \& Jaffe 1984), and we take $\sigma=300$ MeV in this paper (Note, there is a typo in Na et al. (2012) before the term $3 m_{s}^{4}\textrm{ln}^{2}\frac{m_{s}}{\mu_{s}}$ of $\Omega_{s}$, it should be  "$-$" as given by Alcock et al. (1986)).

\begin{figure*}
\centering
\resizebox{\hsize}{!}{\includegraphics{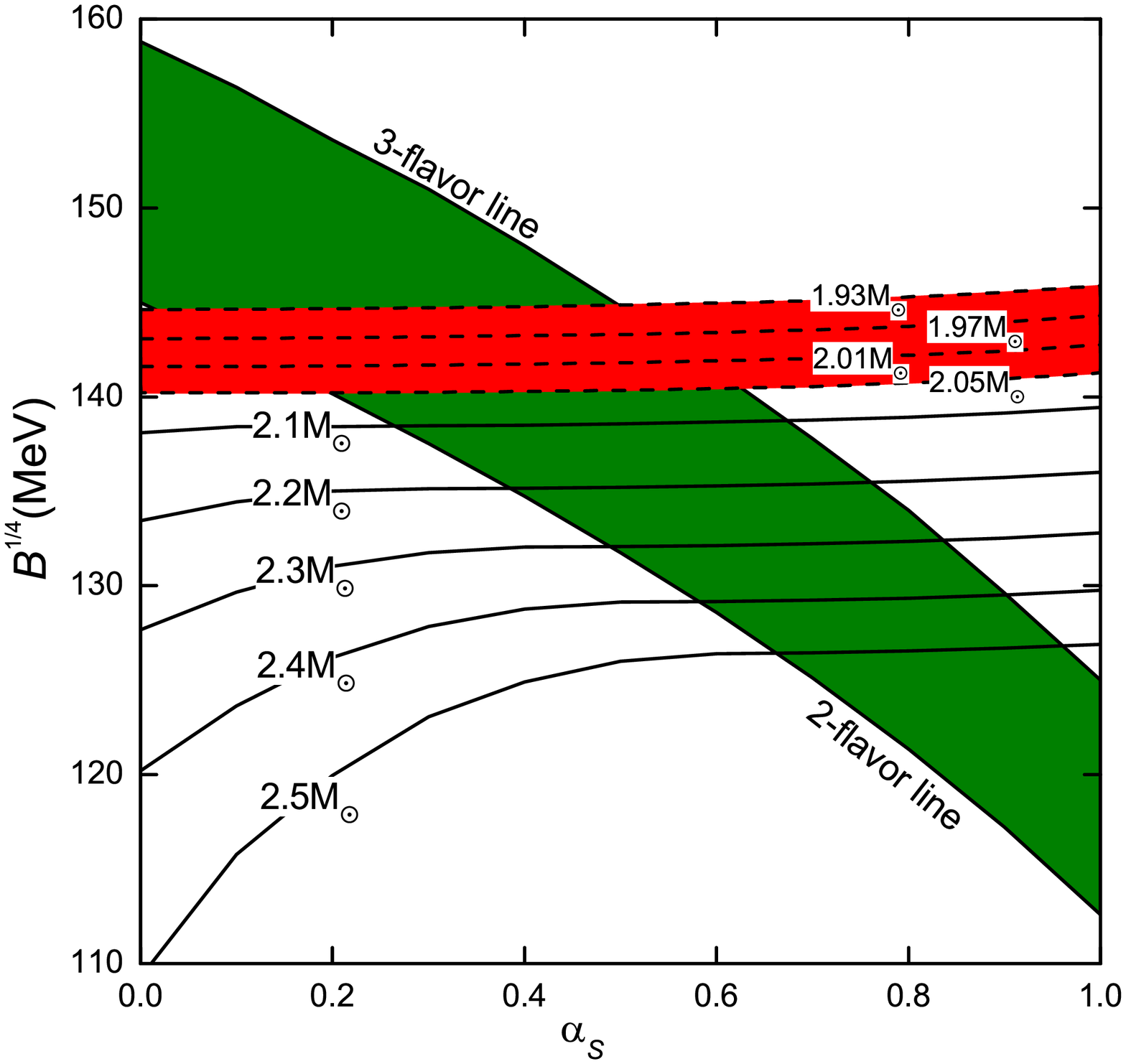}\includegraphics{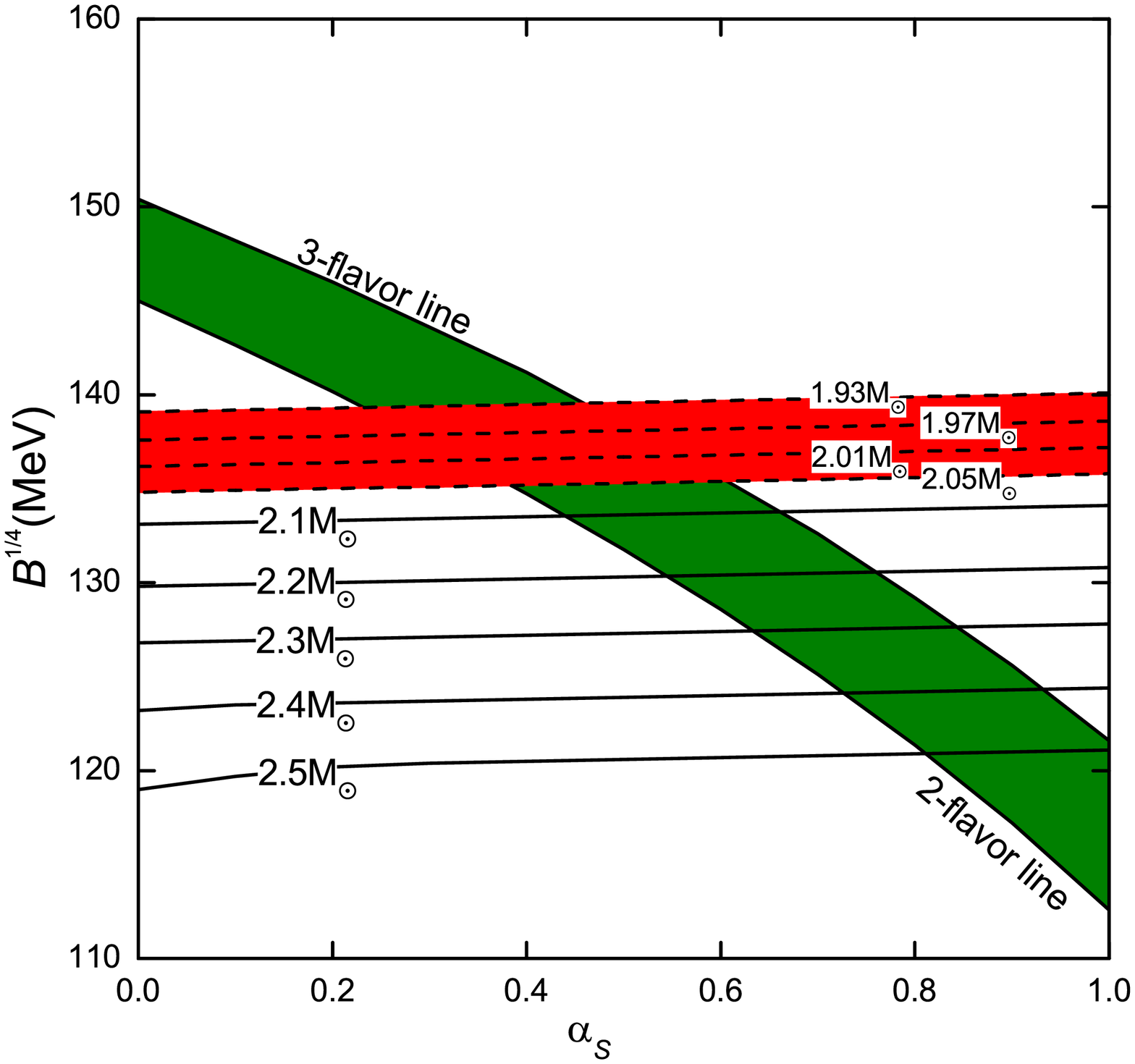}}
         \caption{The constraints to the parameters of the EOS of strange quark matter, namely, $B^{1/4}$ and $\alpha_{S}$. The green shaded area corresponds to the allowed parameter space according to the constraints of the absolute stability of strange quark matter (3-flavor line) and the existence of nuclei (2-flavor line). The red shaded area marks the parameter space which have the maximum mass as PSR J1614-2230 ($M=1.97\pm0.04$ $M_{\odot}$) and PSR J0348+0432 ($M=2.01 \pm 0.04$ $M_{\odot}$). The combinations of $B^{1/4}$ and $\alpha_{S}$ which could lead to the strange star maximum mass as $M$ =2.1 $M_{\odot}$, 2.2 $M_{\odot}$, 2.3 $M_{\odot}$, 2.4 $M_{\odot}$, 2.5 $M_{\odot}$ are also presented, separately. The two graphs are for $m_{s}=100$ MeV (left panel) and $m_{s}=200$ MeV (right panel), respectively.
          }
   \label{fig:1}
\end{figure*}

Before the discussion of the $r$-mode instability window of strange stars and compare it with observations of neutron stars in LMXBs,
we calculate the allowed parameter space of EOS of strange quark matter according to the following basic constraints (Schaab et al. 1997; Weissenborn et al. 2011; Wei \& Zheng 2012). First, the existence of quark stars
composed of stable strange quark matter is based on the idea that
the presence of strange quarks can lower the energy per baryon of
the mixture of $u$, $d$ and $s$ quarks in beta equilibrium below the one
of $^{56}$Fe ($E/A\sim 930$ MeV) (Witten 1984). This constraint results in the "3-flavor line" in Fig. 1.
The second constraint is given by the assuming that non-strange quark
matter (two-flavor quark matter consists of only $u$ and $d$ quarks) in bulk has a binding energy per baryon higher than the
one of the most stable atomic nucleus, $^{56}$Fe, plus a 4 MeV correction
coming from surface effects (Farhi \& Jaffe 1984). By imposing that $E/A\geq 934$ MeV for non-strange quark matter,
one ensures that atomic nuclei do not dissolve into their constituent quarks and gives the "2-flavor line" in Fig. 1.
The last constraint is that the maximum mass must be greater than the masses of PSR J1614-2230 ($M=1.97\pm0.04$ $M_{\odot}$) and PSR J0348+0432 ($M=2.01 \pm 0.04$ $M_{\odot}$).
This constraint can also be shown in Fig. 1, since for each set of parameters of the strange quark matter EOS (namely, $m_{s}$, $B^{1/4}$ and $\alpha_{S}$), one can
give a maximum mass by solving the Oppenheimer-Volkoff equations.

According to the above three constraints, the allowed  parameter space can be decided in Fig. 1.
The region between the "3-flavor line" and "2-flavor line" (the green shaded area) is allowed according to the first two constraints,
but considering the third constraint, only a part of the green shaded area are allowed, namely, the part below the red shaded area. From Fig. 1,
It could be found that for our EOS model, both for the cases of  $m_{s}=100$ MeV and $m_{s}=200$ MeV, the constraint about the maximum mass results in
$\alpha_{S}>0$, which means the QCD corrections must be included, and it is the same result as given by Weissenborn et al. (2011).

\section{$R$-mode instability window of the strange stars}
The $r$-mode instability window of a strange star is decided by the inequality
\begin{equation}
\frac{1}{\tau_{GW}}+\frac{1}{\tau_{\eta}}+\frac{1}{\tau_{\zeta}}<0,  
\label{ineq}
\end{equation}
where $\tau_{GW}$ is the time scale of the growth of an $r$-mode due to the emission of gravitational waves, $\tau_{\eta}$ and $\tau_{\zeta}$
are the dissipation time scales due to shear viscosity and bulk viscosity, respectively. For a strange star with given spin frequency $\Omega$
and core temperature $T$ which satisfy the above inequality, the $r$-mode in the star should increase exponentially, and the amplified
$r$-mode will transmit angular momentum of the star to gravitational waves; therefore, the star should quickly leave the instability window, making vanishing
small probability to observe it in that region in the $\Omega-T$ plane (Gusakov et al. 2014).

The growth time scale due to the emission of gravitational waves is given by Lindblom et al. (1998)
\begin{equation}
\frac{1}{\tau_{GW}}=-\frac{32\pi G \Omega^{2l+2}}{c^{2l+3}}
\frac{(l-1)^{2l}}{[(2l+1)!!]^{2}}\bigg(\frac{l+2}{l+1}\bigg)^{2l+2}
\int_{0}^{R}\rho r^{2l+2}dr,
\end{equation}
where $\Omega$ is the spin frequency of the star, $\rho$ is the mass density in g cm$^{-3}$. In this paper, we only focus on the
$r$-modes with quantum number $l=2$ and azimuthal projection $m=2$ because these are the dominant ones (Lindblom et al.1998; Madsen 1998).

The dissipation time scale due to shear viscosity is (Lindblom et al.1998)
\begin{equation}
\frac{1}{\tau_{\eta}}=(l-1)(2l+1)\bigg[\int_{0}^{R}\rho r^{2l+2}dr\bigg]^{-1}\int_{0}^{R}\eta r^{2l}dr.
\end{equation}
The shear viscosity of strange quark matter due to quark scattering was calculated by Heiselberg \& Pethick (1993), and the results for $T\ll\mu$, where $T$ is the temperature and $\mu$ is the quark chemical potential, can be presented as (Madsen 1998)
\begin{equation}
\eta \approx 1.7 \times 10^{18}\bigg(\frac{0.1}{\alpha_{S}}\bigg)^{5/3}\rho_{15}^{14/9}T_{9}^{-5/3} \mathrm{g\, cm^{-1}\, s^{-1}},
\end{equation}
where $T_{9}\equiv T/10^{9}$ K, and $\rho_{15}\equiv\rho/10^{15}$ g cm$^{-3}$.

The dissipation time scale due to bulk viscosity is given by Refs. (Lindblom \& Owen 2002; Nayyar \& Owen 2006; Vida\~{n}a 2012), considering the second order effects (Lindblom et al. 1999)
\begin{equation}
\frac{1}{\tau_{\zeta}}=\frac{4\pi}{690}\bigg(\frac{\Omega^{2}}{\pi G \bar{\rho}}\bigg)^{2}R^{2l-2}\bigg[\int_{0}^{R}\rho r^{2l+2}dr \bigg]^{-1}
\int_{0}^{R}\zeta \bigg(\frac{r}{R}\bigg)^{6}\bigg[1+0.86\bigg(\frac{r}{R}\bigg)^{2} \bigg]r^{2}dr,
\end{equation}
where $\bar{\rho}\equiv M/(4\pi R^{3}/3)$ is the average density of the nonrotating star. The bulk viscosity of strange quark matter depends mainly on the rate of the non-leptonic weak interaction (Wang \& Lu 1984; Sawyer 1989; Madsen 1992)
\begin{equation}
u+d \leftrightarrow s+u,
\end{equation}
to good approximation the bulk viscosity is (Madsen 1992)
\begin{equation}
\zeta \approx \alpha T^{2}/[\omega^{2}+\beta T^{4}],
\end{equation}
with $\alpha$ and $\beta$ given by Madsen (1992), and $\omega$ is the angular frequency of the perturbation. During the study of $r$-mode instability, $\omega$ is the angular frequency of the $r$-mode perturbation $\omega_{r}=2m\Omega/l(l+1)$, where $\Omega$ is the spin frequency of the star. For the dominant $r$-mode ($m=l=2$), $\omega=\frac{2}{3}\Omega$.
The low-$T$ limit ($T<10^{9}$ K) is enough for this work, and it turns out to be (Madsen 2000)
\begin{equation}
\zeta \approx 3.2\times10^{28}m_{100}^{4}\rho_{15} T_{9}^{2}\omega^{-2} \mathrm{g\, cm^{-1}\, s^{-1}},
\end{equation}
where $m_{100}$ is the strange quark mass in units of 100 MeV and all the other quantities are in cgs units.

\section{Comparison of the instability window with observations }

By solving the inequality (5), together with Eqs. (6), (7) and (9) numerically  for given parameter sets of EOS of strange quark matter, one can get the $r$-mode instability window of strange stars. Here, we want to stress that we will only discuss the parameter sets of strange quark matter EOS which reside in the allowed parameter space as shown in Section 2.

\begin{figure*}
\centering
\resizebox{\hsize}{!}{\includegraphics{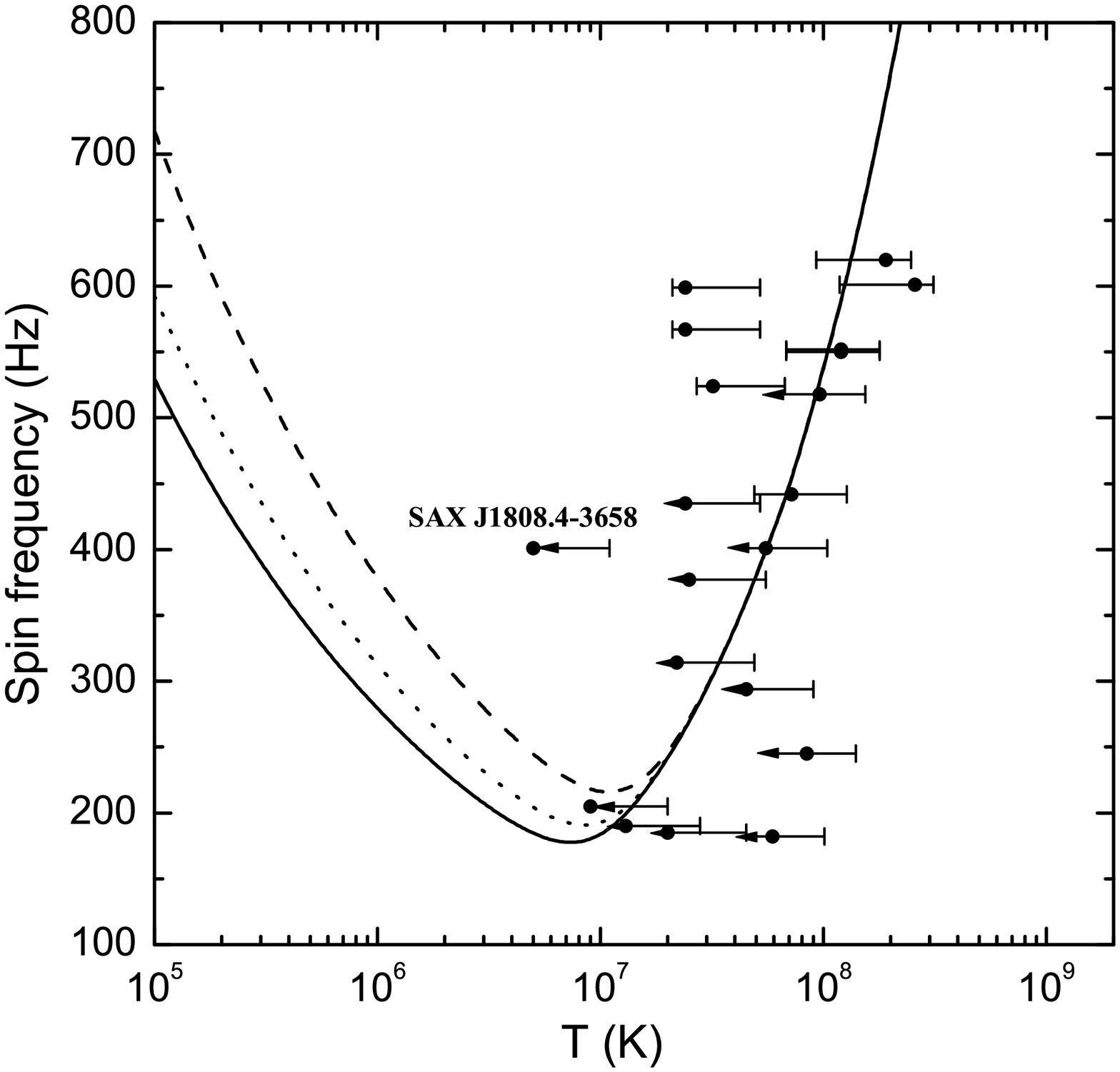}\includegraphics{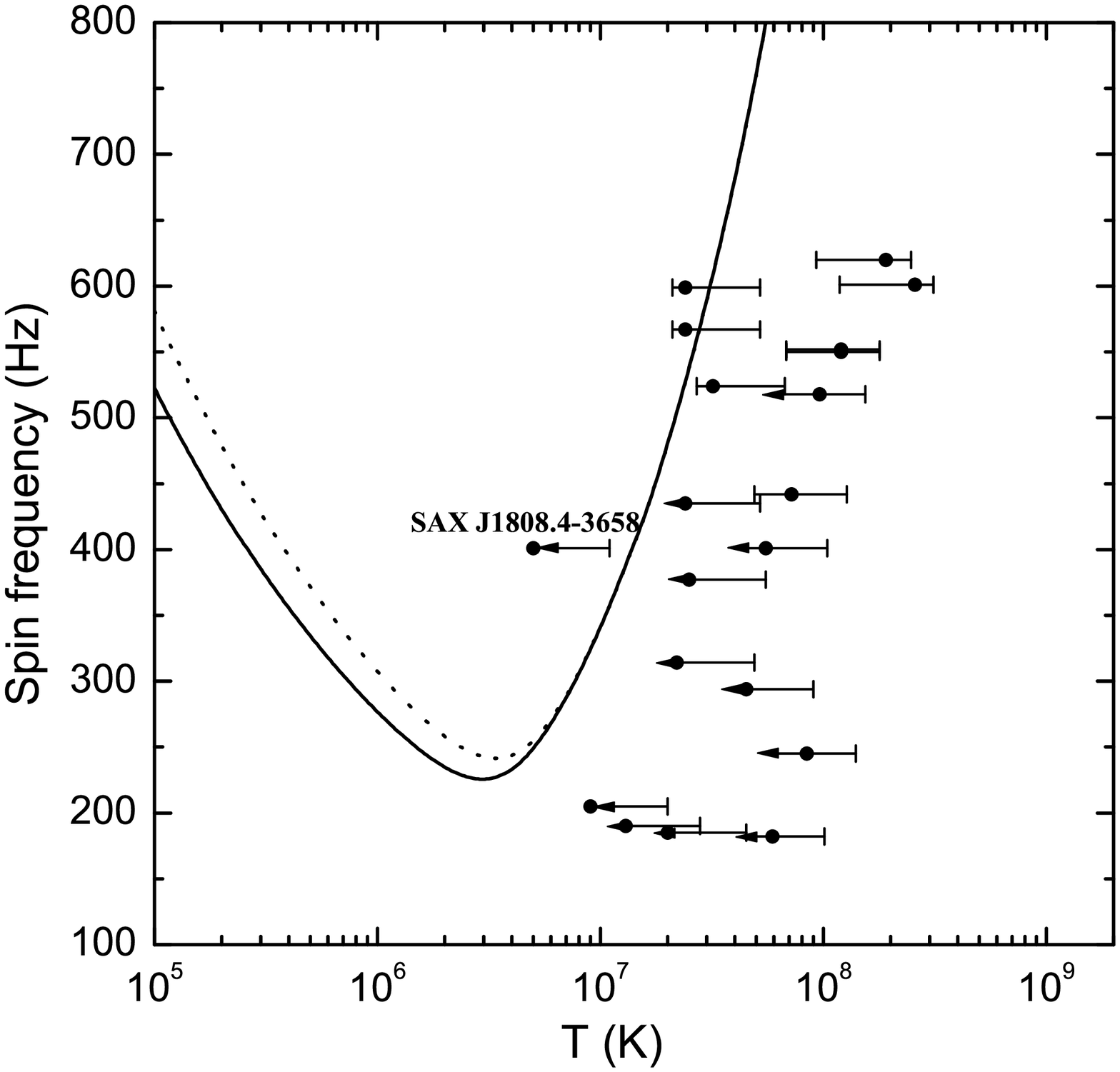}}
         \caption{R-mode instability window for strange star with $M=1.4$ $M_{\odot}$, comparing with the observational data on the spin frequency and internal temperature of neutron stars
          in LMXBs (Gusakov et al. 2014). The left panel is for $m_{s}=100$ MeV and $B^{1/4}=140$ MeV, and the right panel is for $m_{s}=200$ MeV and $B^{1/4}=135$ MeV. The dashed, dotted and solid curves correspond to $\alpha_{S}=0.2$, $\alpha_{S}=0.4$ and $\alpha_{S}=0.6$, respectively(Note, there is no dashed curve in the right panel because non-strange quark matter does not satisfy the condition $E/A\geq 934$ MeV for the parameter set $\alpha_{S}=0.2$, $m_{s}=200$ MeV and $B^{1/4}=135$ MeV, which can be seen in Fig. 1).
          }
   \label{fig:2}
\end{figure*}

\begin{figure*}
\centering
\resizebox{\hsize}{!}{\includegraphics{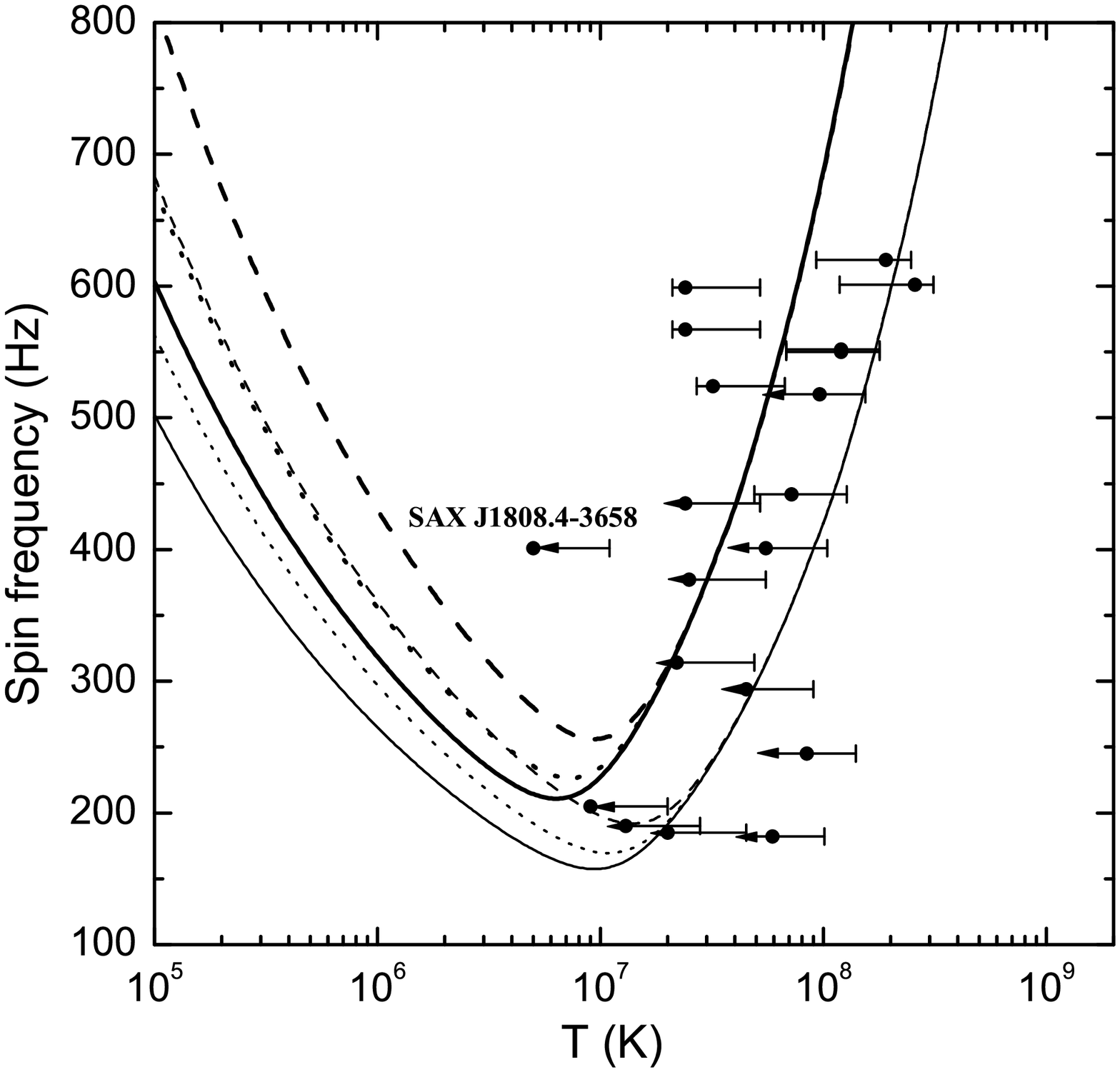}\includegraphics{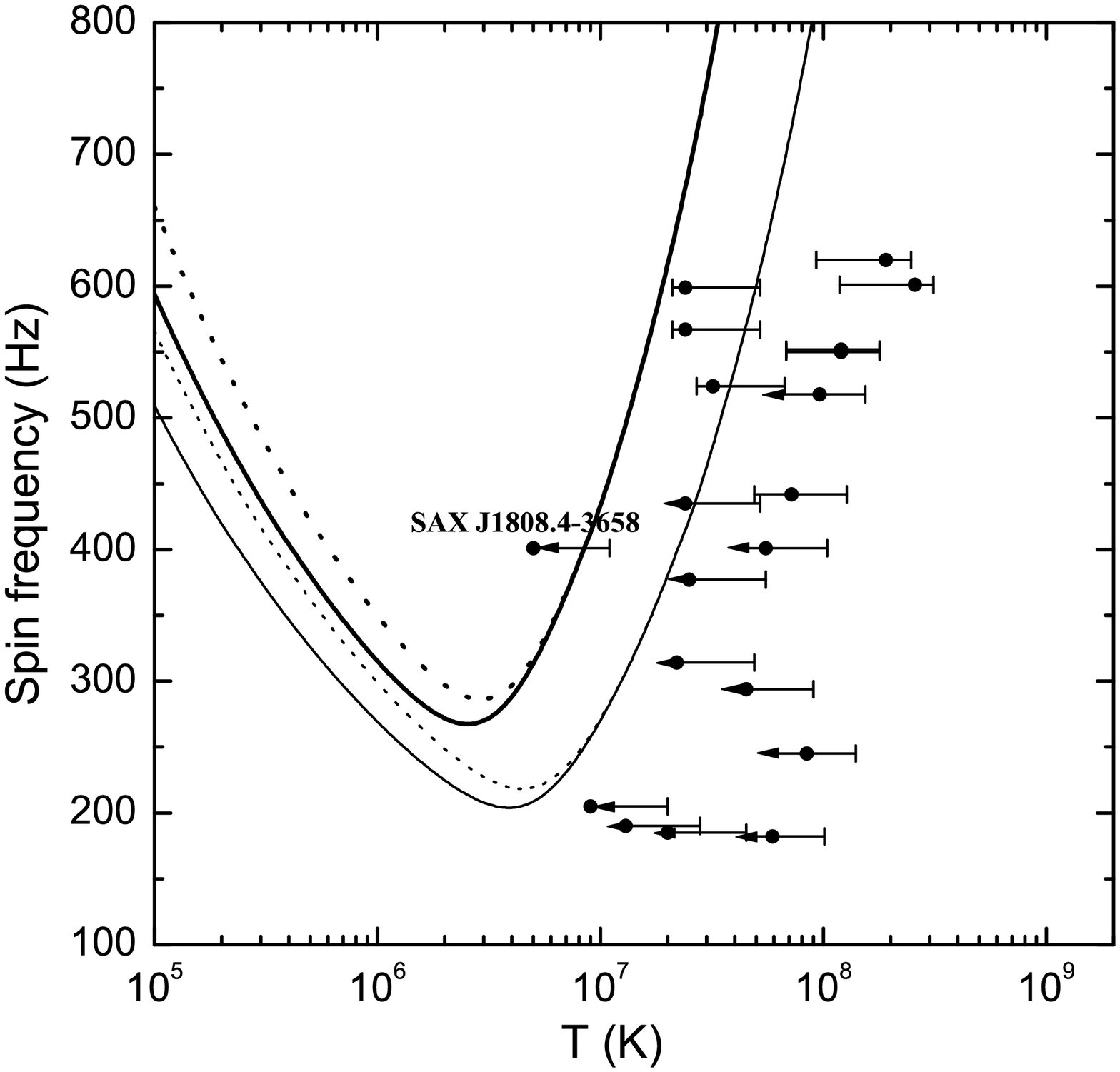}}
         \caption{Similar to Fig.2, but for strange stars with $M=1.0$ $M_{\odot}$ (thick lines) and $M=2.0$ $M_{\odot}$ (thin lines).
          }
   \label{fig:3}
\end{figure*}

Fig. 2 shows the $r$-mode instability window for strange star with the canonical neutron star mass $M=1.4$ $M_{\odot}$, and the observational data on the spin frequency and internal
temperature of neutron stars in LMXBs, which are given by Gusakov et al. (2014). The left panel is for $m_{s}=100$ MeV and $B^{1/4}=140$ MeV, and the right panel is for $m_{s}=200$ MeV and $B^{1/4}=135$ MeV (For each given $m_{s}$, we select the nearly largest $B^{1/4}$ that allowed by the limit of observational neutron star mass according to Fig. 1, because it corresponds to smaller allowed $\alpha_{S}$, which will lead to smaller $r$-mode instability region as can be seen in Fig. 2). For the left panel, three curves are presented, which represent the case of  $\alpha_{S}=0.2$, $\alpha_{S}=0.4$ and $\alpha_{S}=0.6$, respectively; while for the right panel, we only show two curves, namely, $\alpha_{S}=0.4$ and $\alpha_{S}=0.6$. The reason is that the parameter set $\alpha_{S}=0.2$, $m_{s}=200$ MeV and $B^{1/4}=135$ MeV is not located in the allowed parameter space as discussed in Sect. II, more exactly, non-strange quark matter does not satisfy the condition $E/A\geq 934$ MeV for this parameter set. It could be seen that, all the possible instability windows are not consistent with the spin frequency and temperature observations of neutron stars in LMXBs, which turns out to be the same conclusion as drawn by Haskell et al. (2012).

In Fig.3, we present the $r$-mode instability window or strange stars with  $M=1.0$ $M_{\odot}$ and $M=2.0$ $M_{\odot}$. From this graph, one can draw the same conclusion as the above one from Fig.2.  However, as one can seen form the thick curves of the right panel of Fig. 3, for fixed parameter set $M=1.0$ $M_{\odot}$, $m_{s}=200$ MeV and $B^{1/4}=135$ MeV, both of the two curves (for $\alpha_{S}=0.4$ and $\alpha_{S}=0.6$, respectively) can well explain all the observational data except that for SAX J1808.4-3658. Therefore, the soure SAX J1808.4-3658 is very important to us when drawing any conclusions.

\section{Conclusion and discussion}

Following the brief study by Haskell et al. (2012), we examine the instability window of strange stars in detail, and compare it with the spin frequency and temperature observations of neutron stars in LMXBs. Our work is based on a realistic EOS of strange quark matter, namely, the modified bag model. Besides the numerical calculation to the timescales related to $r$-modes, we also employ a delicate strategy, in which firstly, we calculate the allowed parameter space of EOS so that it can match the observed mass limit of neutron stars, and then the study of the instability window of strange stars and its comparison with the observations are carried out.

\begin{figure*}
\centering
\resizebox{\hsize}{!}{\includegraphics{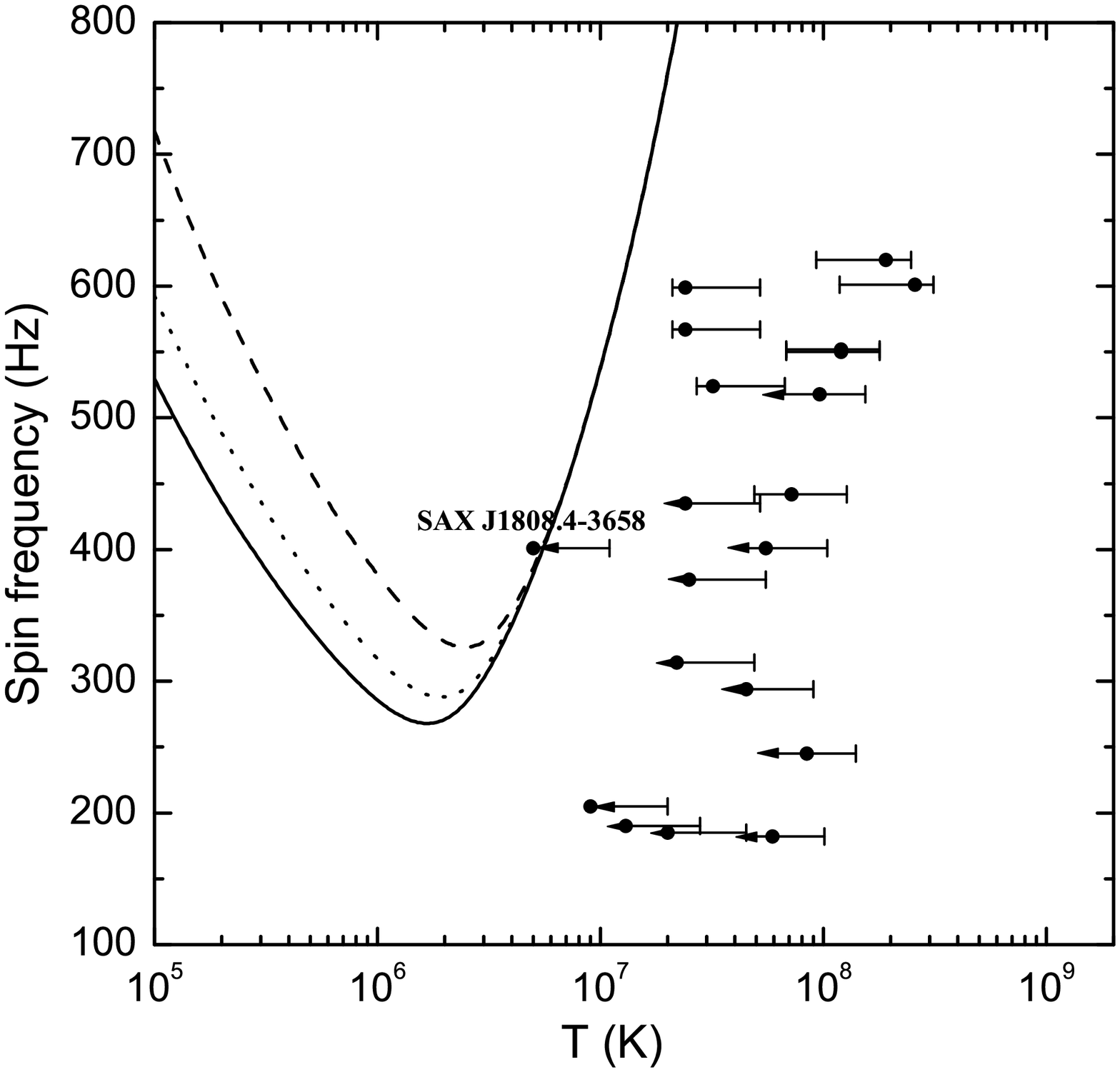}\includegraphics{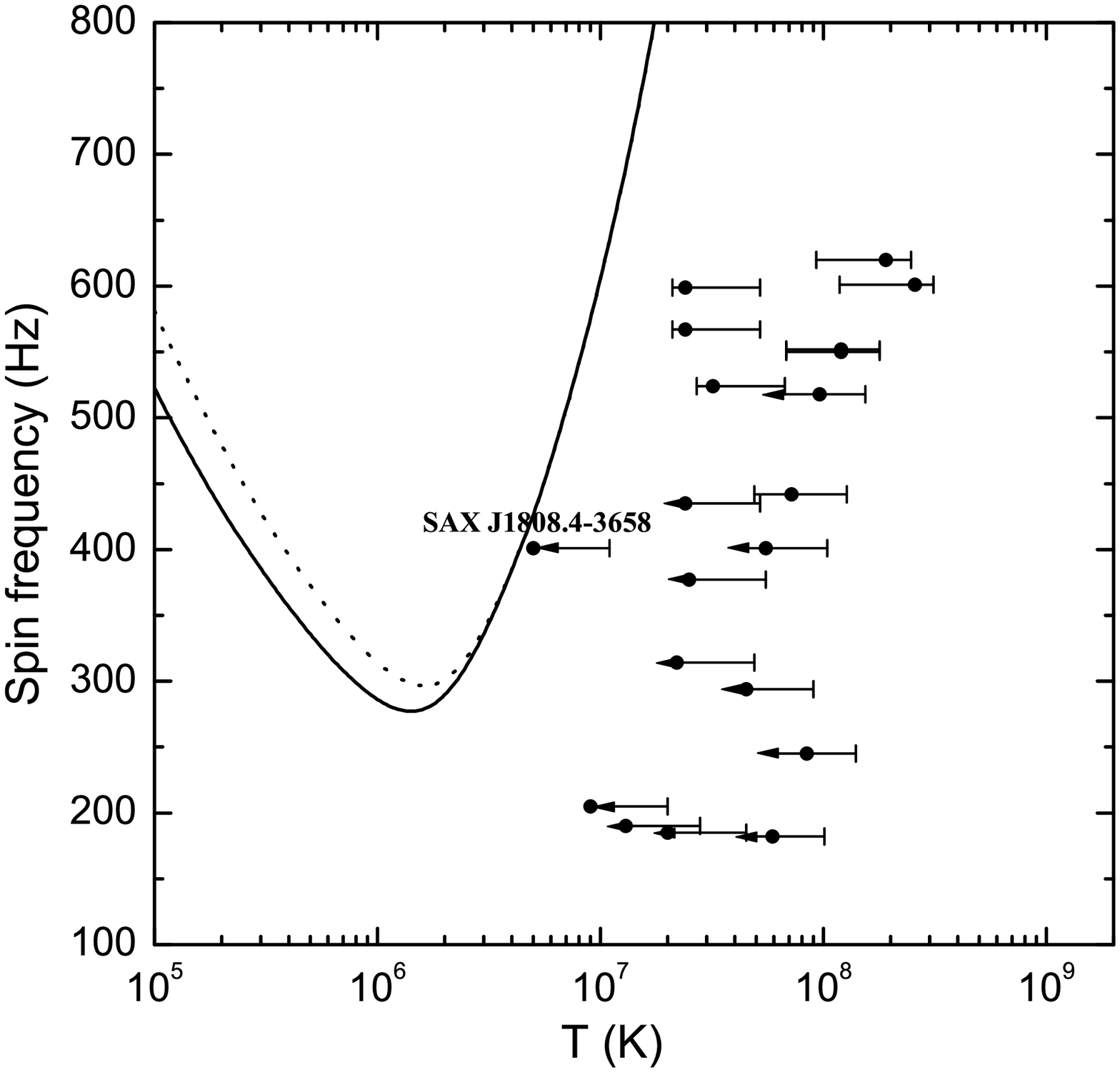}}
         \caption{Similar to Fig.2, but $\zeta$ is artificially taken as a 100 times larger one for the left panel and a 10 times larger one for the right panel.
          }
   \label{fig:3}
\end{figure*}

Our study confirms the conclusion given by Haskell et al. (2012) that all the possible instability windows of strange stars are not consistent with the spin frequency and temperature observations of neutron stars in LMXBs. However, as far as the bulk viscosity of strange quark matter is considered in this paper, it is calculated under non-interacted Fermi liquid model (Madsen 1992). If the interactions which lead to non-Fermi liquid effects are included, the bulk viscosity $\zeta$ can be increased by many orders of magnitude (Zheng et al. 2002; Zheng et al. 2003; Zheng et al. 2005; Schwenzer 2012), and the instability window may be consistent with the observations. This possibility is shown roughly in Fig. 4, using the same parameter sets of EOS as Fig. 2 but the bulk viscosity $\zeta$ is artificially taken as a 100 times larger one for the left panel and a 10 times larger one for the right panel. It could be seen from Fig. 4 that the instability window could almost be consistent with the observations under the above assumptions. The detailed study about that possibility will be carried out in our future work.

\section*{Acknowledgments}
The authors want to thank the anonymous referee for his/her kindly suggestions. One of the authors, S. H. Yang is grateful to F. Weber for useful discussions related to this work.
C. M. Pi is supported by the Scientific Research Program of Hubei Provincial Department
of Education (No.Q20123101) and the CCNU-QLPL Innovation Fund (QLPL2013P01). S. H. Yang is supported by the National Natural Science Foundation of
China (NSFC) under Grant No.11203010 and the college’s
basic research and operation of MOE of China (Grant No. CCNU13A05038).
X. P. Zheng is supported by the Key Program Project of Joint Fund of Astronomy by NSFC and Chinese Academy of Sciences (No.11178001).

\end{document}